\documentstyle[epsf,epsf]{aipproc}
\begin{document}
\title{Star Formation History of Elliptical Galaxies
from Low-Redshift Evidence}

\author{Guy Worthey\thanks{Hubble Fellow} }
\address{Astronomy Department, University of Michigan \\
Ann Arbor, MI 48109-1090}

\maketitle

\begin{abstract}
Star formation in elliptical galaxies (Es) was and is mostly dominated
by mergers and accretions with many suggestive examples seen among local
galaxies.  Present day star formation in Es is easily measurable in
\twothirds\ of Es and appears bursty in character. Direct age
determinations from integrated light indicate real age scatter. If one
assumes the oldest-looking galaxies are a Hubble time old, the light
weighted mean ages of the rest spread to 0.5 of a Hubble time, with
scatterlings at very young ages. Larger Es and Es in clusters have
less age scatter than smaller or field Es. The size trend is
clear. The environment trend needs to be rechecked with better data
even though it agrees with high redshift field/cluster results.
\end{abstract}

\section*{Introduction}

The most appealing picture of star formation in elliptical galaxies for
cosmologists is one in which Es formed very early in the universe and
have been quiescent ever since. If such galaxies exist, they
potentially measure the curvature parameter $q_0$ since one could
predict their size and luminosity fairly well and use them as
standard rulers or candles. This picture is supported by the
superficial uniformity of ellipticals in appearance and the
existence of scaling relations between observed parameters (stellar
velocity dispersion [$\sigma$], surface brightness, size, luminosity,
colors, and line strengths), some of which scale with very small
scatter.

The hope for using Es as cosmological tracers is alive and well, but when
nearby ellipticals are examined in detail they exhibit a large variety
of morphological and kinematic peculiarities and strong evidence for
star formation much less than a Hubble time ago, leading to a picture
of elliptical formation in which galaxy-galaxy mergers and accretion
events are the dominant formation mechanism, and that we still witness
the ``tail end'' of E formation today in the form of observed galaxy
mergers and Es that display current star formation. I think we should
view elliptical formation as a {\it process} rather than an {\it event}.

\begin{figure} 
\epsfysize=3.0in
\centerline{\epsfbox[35 406 576 750]{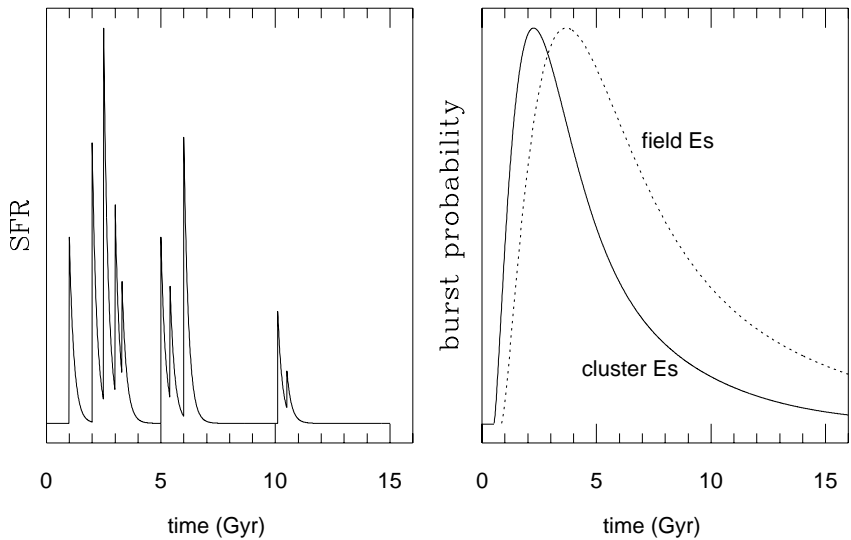}}
\caption{ A schematic picture of the most probable way that star
formation proceeded in E galaxies over the life of the universe. Each
galaxy is formed by a series of accretions and mergers.  The strengths
and times of the star formation bursts are random and have occurred in
precursor branches of a merging tree. Bursts are governed by an
overall probability distribution in which star formation is more
likely to have occurred in the past. The probability curve 
varies as a function of field/cluster environment and galaxy size.  }
\vspace*{10pt}
\label{fig1}
\end{figure}

Figure 1 does not show the variety of dynamical or morphological
changes that could be taking place in the history of an elliptical,
but concentrating on star formation alone, a bursting scheme
like that of Fig. 1 fits the facts outlined in this article. Note in
Fig. 1 a difference in mass-weighted age between cluster
and field ellipticals and the bursty star formation with quiescent
periods during which ellipticals will look superficially normal if the
intraburst period is longer than the stellar population fade time.

This fade time is illustrated in Fig. 2 using the commonly used and
easily measured line index Mg$_2$, which has the additional advantage
that it participates in the Mg$_2$-$\sigma$ scaling relation (discussed
later) so that there is a relatively clear definition of ``normal:''
the gaussian scatter from this relation. Fig. 2 shows that, if
only a few percent by mass of gas is consumed in a starburst, the
elliptical galaxy will appear ``normal'' after less than 0.5
Gyr. Therefore a population of Es will always look ``normal'' except
for a few oddballs caught in the aftermath of starburst.
The rest of this article should show that this is a workable scheme.

\begin{figure} 
\epsfysize=2.5in
\centerline{\epsfbox[35 396 576 750]{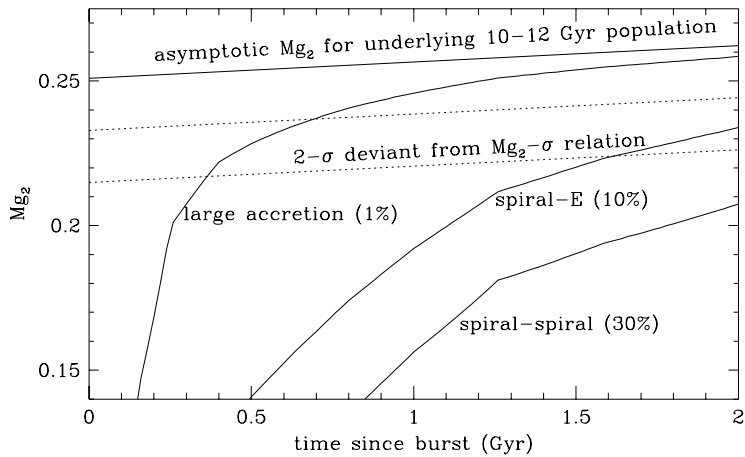}}
\caption{ 
The
Mg$_2$ evolution of different fractions by mass (1\%, 10\%, and 30\%)
of burst mass / total mass are shown to
illustrate how fast an elliptical galaxy
will look ``normal'' after an episode of star formation.
Time is counted starting from a 
burst of star formation that occurred in a passively evolving, 10-Gyr-old
``elliptical,'' whose unperturbed Mg$_2$ strength is also shown.  
One-$\sigma$ and 2-$\sigma$
increments from the Mg$_2$-$\sigma$ relation ($\sigma_{\rm measurement}=
\pm 0.018$ mag [Bender et al. 1993])
are shown to indicate when a
galaxy would ``blend in'' with the others. The error in Mg$_2$
is about 0.01 mag. The models shown in this diagram are a
variant of the Worthey (1994) models in which Padova evolution is used
instead of VandenBerg. }
\vspace*{10pt}
\label{fig2}
\end{figure}

\section*{Local evidence for merging}

I will differentiate between ``merging'' and ``accretion''
events. Accretion events are small, involving less than 10\% of the
galaxy's mass, while mergers are larger. Each event can also be
partially or completely stellar, so that star formation may not
result. Theoretical expectations are that gaseous accretions will
rapidly, dissipatively sink to galaxy center, forming stars when it
becomes dense enough. Stellar accretions survive as kinematic substructure
that disappears if there is time to phase mix (e.g. Kormendy
1984; Balcells \& Quinn 1990; Balcells 1991). Large, violent mergers
are characterized by the ejection of tidal tails and a somewhat
protracted series of starbursts as the two galaxies violently relax to
an $r^{1/4}$ profile, followed by a steady drizzle of gas that can
last as long as a few Gyr (Hibbard \& Mihos 1995). The end state of
such violent encounters has long been postulated to end in the
formation of a bulge-dominated elliptical-like galaxy, with
``correct'' kinematics and surface brightness profile (e.g. Barnes
1992, Hernquist 1993, Heyl et al. 1994) if a little gas is
involved. Note that purely stellar mergers give very large cores that
are not observed, so a certain amount of gaseous dissipation is
required.

{\bf Ongoing disk-disk mergers:\quad} Many IRAS galaxies and Arp
(1966) peculiar galaxies appear to be disk-disk mergers in progress:
still recognizable as two spiral galaxies, but blatantly
interacting. Toomre (1977) counted 11 such mergers with NGC
numbers. Assuming a constant merger rate and a typical duration of 0.5
Gyr for this morphological phase found that $\sim$250 NGC galaxies
would have formed this way over a Hubble time.  Compare this number to
the fraction of the 6032 (my estimate) galaxies in the original NGC
catalog that are Es: about 13\% = 780 objects.

{\bf Qualitative post-merger signatures:\quad}
Both morphological and kinematic peculiarities are predicted outcomes
of large accretions or mergers.
Morphological ``fine structure,'' including
ripples, jet features, boxiness, and X structure at large radius was
measured by Schweizer et al. (1990) and Schweizer \& Seitzer
(1992). The galaxies with the most such morphological complexity have
systematically bluer colors, weaker metallic line strengths, and
stronger H$\beta$ line strength, indicating a connection between
recent merger status and the mean age of the stellar populations (the
other alternative is that morphologically disturbed galaxies are
preferentially metal-poor, but this idea is generally
dismissed).  

Kinematic peculiarities such as minor axis rotation, cores that rotate
counter to the rotation of the galaxy outskirts, and kinematic
discontinuities with radius are observed in more than half of all
ellipticals (Bender 1996). Many of these peculiarities are dynamically
long-lived, and do not constitute evidence for or against {\it recent}
merging, but do argue strongly for merging as a process, since a
single radial collapse can not produce such peculiar
motions. Kinematic discontinuities are often mirrored by cospatial line
strength gradient discontinuities, adding further weight to the merger
interpretation (Bender \& Surma 1992). In the case of high surface
brightness elliptical NGC 1700, Statler et al. (1996) argue from an
array of isophotal shape and kinematic data that at least 3 or more
stellar subsystems must have merged 2--4 Gyr ago to explain the
observed substructure, but the inner regions have phase-mixed to
uniformity. Very few galaxies have been studied to this level of detail.

{\bf Accretion in Non-Es:\quad}
A number of galaxies illustrate that multi-episode gaseous
accretion/merger events occur. A note of thanks to Kennicutt (1996)
for pointing out most of the examples I describe. First, we catch an
accretion event in progress in the Milky Way as the Saggitarius dwarf
spheroidal is in the process of being tidally disrupted (Ibata et
al. 1995) and is strung out over at least $\sim$40 degrees of arc
(M. Mateo, private communication). More evidence for the accretion of
spheroidal-sized, metal-poor, but not always old galaxies into the
Galactic halo comes from young stars on halo orbits, the presence of
galactic streams, and age spreads in globular clusters as summarized
in Freeman (1996). Some workers now start with the assumption that the
entire metal poor halo was built by accretion events over the Galaxy's
lifetime, while the bulge may have been built from disk/bar
instabilities. It does seem likely that bulges and disks evolved
together since their scale lengths are always in nearly the same
ratio (Courteau et al. 1996).

Accretion of gas is observed in external spiral galaxies from H {\sc
I} in the form of high velocity clouds of up to $\sim 10^8 M_\odot$
(Kamphius 1993), or in the form of blatant tidal disruption prior to
merging as in the case of NGC 3359 and NGC 4565 (Sancisi et al. 1990).
H{\sc I} observations of E galaxies NGC 4472, UGC 7636, NGC 3656, NGC
5128, and NGC 2865, among others, show evidence for ongoing or recent
gas accretion (Sancisi 1996). Spiral NGC 4826 (M64) is a rare example
of gas-gas counter-rotation in which the sense of rotation switches at
a radius of $\sim$1 kpc. This galaxy appears to have accreted a
gas-rich companion in the recent past, with the inner gas disk the
remnant of the original disk
(Braun et al. 1992; Rubin 1994; Rix et al. 1995).

Many star-gas counter-rotations are seen. Bertola et al. (1992) list 9
examples of S0 galaxies with gas-star counterrotation. In NGC 4526
(Bettoni et al. 1991) and NGC 3626 (Cirri et al. 1995) the
counterrotation extends over the entire disk, with gas masses of $10^8
- 10^9 M_\odot $. More easily seen are polar ring galaxies in which a
(usually) spiral galaxy is ringed by gas, stars, and/or dust at a
nearly perpendicular angle. From their catalog of ($\sim$70) polar
ring galaxies, Whitmore et al. (1990) estimate that about 5\% of S0
galaxies went through a polar ring phase.
Even more spectacular than gas-star is star-star
counterrotation. The edge-on S0 galaxy NGC 4550 is seen to have two
cospatial counterrotating disks with nearly identical masses and scale
lengths and with velocity dispersions of 45 and 50 km s$^{-1}$ (Rubin
et al. 1992; Rix et al. 1992), indicating two separate epochs of
galaxy formation. In Sb NGC 7217 20-30\% of the the disk
stars are in a retrograde cold disk (Merrifield \& Kuijken 1994). In
spirals, such oddities are rare probably because spirals are
relatively fragile compared to Es. 

The evidence from spirals indicates that accretion is a
common phenomenon. Further, although galaxies like NGC 4550 must be
rare, they reveal a fabulous wealth of possibilities for galaxy
formation. Before you knew about 4550, what odds would you have
assigned to its existence? 

{\bf Current E star formation:\quad}
Most Es have readily measurable nebular emission (Goudfrooij et al. 1994,
Gonz\`alez 1993) probably indicative of star formation. The Goudfrooij
narrow band imaging indicates a variety of morphologies of the ionized
gas, from disk-like to nuclear to diffuse and filamentary. 
Fig. 3 shows
reprocessed Gonz\`alez (1993) spectroscopic data on Es and 
Kennicutt (1983) H$\alpha$ data on spiral galaxies with the same
conversion from H$\alpha$ flux to star formation rate.  Fig. 3
shows the fraction of the galaxy assembled over 10 Gyr assuming a
constant star formation rate. I
corrected roughly for $M/L$ changes as a function of Hubble type, but
assumed zero extinction correction. A correction will push
the star formation rate to higher values.

\begin{figure} 
\epsfysize=3.0in
\centerline{\epsfbox[35 336 576 750]{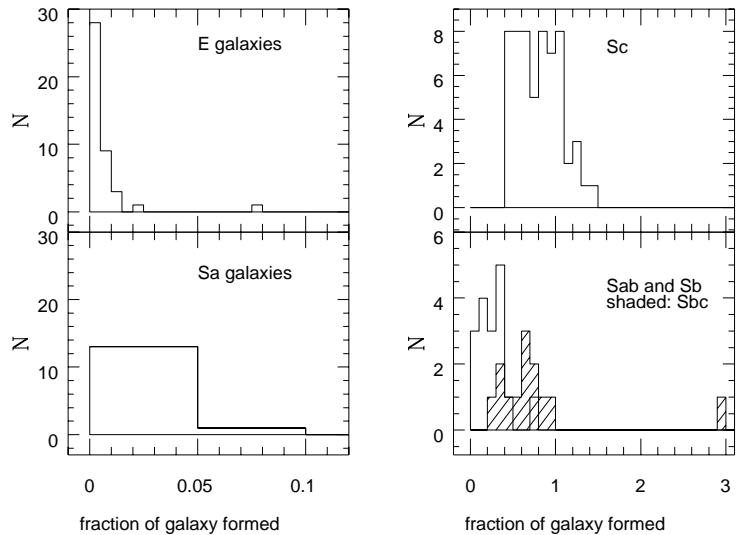}}
\caption{ Current star formation rates for galaxies of different
Hubble types expressed as the fraction of mass that would be assembled
over 10 Gyr assuming the current SFR stayed constant.
Kennicutt (1983) H$\alpha$ and Gonz\`alez (1993) O[III] 
data were used to estimate the SFR via the Kennicutt (1983)
formula with no correction for extinction. The binning of
the Sa data reflects the larger uncertainty in Kennicutt's data.
}
\vspace*{10pt}
\label{fig3}
\end{figure}

There are two points to notice about the elliptical galaxies in relation
to spirals. (1) The observed SFR is nonzero in \twothirds\ of the
Es, but 100 times smaller than late-type spirals. (2) Fig. 3 might
almost be an illustration of binomial statistics in which Sc
galaxies have large numbers of star formation events and are thus
distributed in an almost gaussian way, while E galaxies have small
numbers of star formation events and are thus distributed in a way
that resembles Poisson statistics: {\it fundamentally bursty}
in character.

\section*{Age from integrated light}

It is now possible to measure a light-weighted mean age using
integrated light indices for old stellar populations. The
``light-weighted'' part means that young stars, because they are
brighter, can heavily influence the mean age that one obtains and
that it is easy to obscure older generations of stars. The mean ages
are derived from plotting Balmer indices versus hand-picked metal
indices that are more metallicity sensitive than average. The spectra
from which these measurements are taken need to have good S/N and need
to have careful accounting for systematics like instrumental
resolution and galaxy velocity dispersion so that the observations
transform to the same system as the models.

Such pickiness is needed because easier measurements, such as
broad-band colors, D4000, or Mg$_2$ are largely degenerate with age
and metallicity along a null spectral change slope of d log($Z$) =
$-2/3$ d log(age) along which colors and most line strengths stay the
same (Worthey 1994). This implies that if one wants a 15\% age estimate, one
must know the metal abundance to 10\%.
By picking spectral indices that are preferentially
sensitive to age and arraying them against indices that are
preferentially sensitive to metal abundance the
age-metal degeneracy is largely broken in a {\it differential} sense:
the age zeropoint is still quite uncertain, but relative mean age
changes are readily detectable.

\begin{figure} 
\epsfysize=3.5in
\centerline{\epsfbox[35 396 576 750]{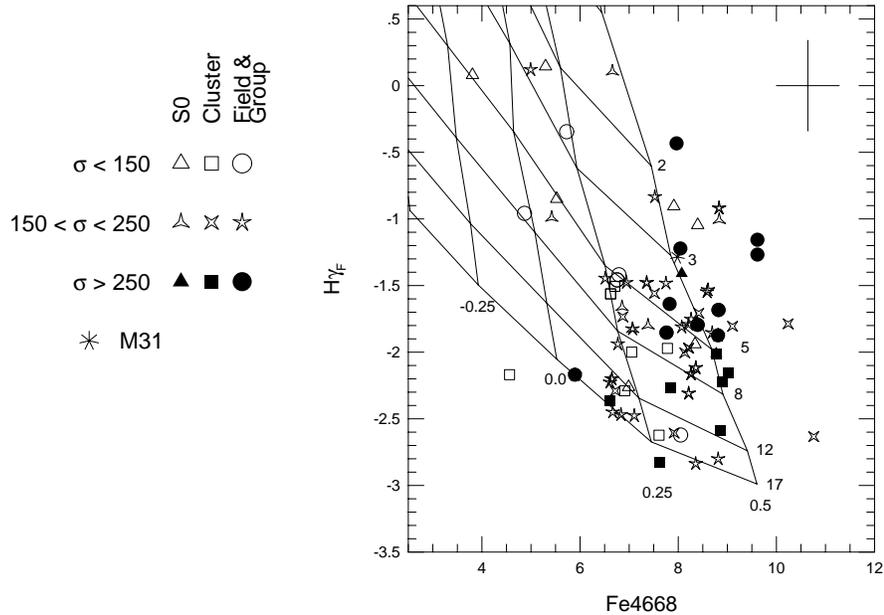}}
\caption{ Early type galaxies of different size and
environment in an age diagnostic diagram; H$\gamma_F$ (Worthey \&
Ottaviani 1997) versus Fe4668 (Worthey et al. 1994).  Models (Worthey
1994) are plotted as a grid labeled with [Fe/H] along the bottom and
age in Gyr along the right edge. Approximate observational error is
indicated.  Three galaxies at very young age and with $\sigma <
150$ km~s$^{-1}$ are off the scale of this diagram.  }
\vspace*{10pt}
\label{fig4}
\end{figure}

Figure 4 tells us several things. First, the spread in mean age is
real in the sense of being well beyond observational error. Regardless
of age zeropoint, many galaxies have ages less than half a Hubble
time, and several appear very young. At present we can't tell if these
ages are 
the real formation ages of the galaxies or the effect of small,
recent bursts of star formation.
Second, there is a distinct trend for the
younger large galaxies to be more metal rich, a tendency which will be
discussed below.  Third, there are dependencies on size and probably
on field/cluster environment. Smaller galaxies have a larger
spread in mean age than the large ones, as well as a tendency to be
slightly more metal poor. Field galaxies appear to be more volatile
than cluster galaxies (note the lack of large cluster Es younger than
3 Gyr). In this sample there do not appear to be any cluster Es that
are both small and young. Note that the sample was not
chosen in an intelligent way, and a proper volume-limited sample may
show somewhat different trends.

The real scatter of ages tells us that Es have had a fairly complex
history, and that they are not quite finished forming. It does
not tell us the relative importance of mergers versus
accretions. 

{\bf Aside: Age from 
abundance ratio variations?\quad}
Worthey et al. (1992) plotted Mg$_2$ versus iron indices
to find that large Es deviated from solar-neighborhood [Mg/Fe] in the
sense of enhanced Mg relative to Fe by about a factor of two. Smaller
Es  have a nearly solar mixture, so the amount of
enrichment from Type II supernovae relative to Type I gets larger in
larger Es. Unfortunately, the mechanism for varying this ratio is not
known. It could either be a variation in formation timescale or a
(mild) variation in upper IMF strength as a function of galaxy
size. The implications for E formation are different for those two
cases and so, skipping the details, we can't really constrain E
formation until we know for sure what mechanism causes the Type II to
Type I shift.

\section*{Slipperiness of tight scaling relations}

As mentioned above, 
colors and line-strengths usually scale with structural parameters
like brightness, size, $\sigma$, and combinations of these
quantities. The tightest are the Mg$_2$-$\sigma$ relation and
the fundamental plane (in $\mu$, $R$, $\sigma$ space).  The interesting
thing about these ``tightness relations'' is the small scatter
observed.
The small scatter is
interesting because significant spread in velocity anisotropy, density
structures, and stellar ages could reasonably be expected to raise the
scatter to much higher levels than is observed (e.g. Djorgovski et
al. 1996). More than one conspiracy must be operating to thus limit
the range of E properties.

Empirically, somewhat more than half the scatter from the fundamental
plane can be fairly unambiguously tied to stellar population changes
because the scatter correlates with color and line strength residuals
such that blue colors correspond to high surface brightness, as one
would expect from the presence of a younger subpopulation (Prugniel \&
Simien 1996; J{\o}rgensen \& Franx 1996). If that is so, then the
tightness of the fundamental plane places a restriction on the age
scatter that is allowed for ellipticals. As a first cut, one can take
the observed scatter in cluster Es and convert
that to a scatter in age or metallicity.  Bender et al. (1993) find,
analyzing the Mg$_2$-$\sigma$ relationship, an allowed age (or
metallicity) scatter of 15\% RMS at a given $\sigma$ with a
non-gaussian blue tail. 
This seems to say that
\twothirds\ of Es were formed in the first \onethird\ of the
universe. This interpretation is probably misleading because (1) the
scatter does not look one-sided -- it looks like there really are
galaxies on the redder (older or more metal rich) side of the mean
relation, and (2) there appears to be a rough trend for younger Es to
be more metal rich (Figure 4; Worthey et al. 1996). Point (1) means
that either metallicity plays a significant/dominant role, or that the
mean age for Es is substantially younger than that of the universe,
and what we are seeing is scatter about a ``mean history'' of
formation.
Point (2) implies that younger (bluer)
populations will be more metal rich (redder), and this will do a lot
to artificially tighten the Mg$_2$-$\sigma$ and the fundamental plane
relations, allowing more age spread than one might otherwise have
guessed.

\section*{Summary and high-redshift remarks}

The tightness relations can also be tracked with redshift. If Es are
pure passive evolvers, there is a clear prediction for bluer Mg$_2$
and brighter $M/L$ with redshift. If we think of E formation as a
decaying process of continued activity, we can predict a lot less
clearly what is going on. In fact, if an ongoing accretion {\it
process} dominates the line strengths and colors, we may see {\it
less} evolution than purely passive. If large bursts of star formation
at fairly late times are important we should see {\it more} 
than passive evolution. Work on both the
fundamental plane and the Mg$_2$ sigma relation in cluster Es out to
redshifts of 0.5 or so (see Dressler, this volume; Bender et
al. 1996) indicates that
the galaxies identified as Es tend to evolve as passively
evolving populations that were formed before $z = 2$.

In the field, however, two different redshift surveys show that
the population of red galaxies declines too fast to be consistent with
passive evolution of old luminous galaxies with extrememly high
confidence (Kauffmann et al. 1997).
These high redshift results lend confidence that the Fig. 1 picture is
roughly correct.
In that picture, field and
cluster Es form by some mixture of merging and accretion 
in a bursty manner which allows lots of time for
blue colors to fade and for the galaxy to look ``normal'' between
bursts. Active star
formation should be seen in some ellipticals. Merging events and
merger remnants should exist. Tight scaling relations are preserved
until star formation gets really messy, probably before $z=2$.
The Balmer-metal age-diagnostic diagrams should show a spread in
age. The Fig. 1 picture seems to summarize the observed situation
fairly well.

\end{document}